\documentclass[prl,twocolumn,amsmath,amssymb,showpacs,floatfix]{revtex4}
\usepackage{graphicx}
\usepackage{bm}
\usepackage{txfonts}

\usepackage{dcolumn}
\newcommand{\ket}[1]{\mbox{$\vert #1 \rangle$}}
\newcommand{\bra}[1]{\mbox{$\langle #1 \vert$}}
\newcommand{\Rb}{\mbox{$^{87}$Rb}}
\newcommand{\up}{\mbox{$\vert \! \uparrow \rangle$}}
\newcommand{\down}{\mbox{$\vert \! \downarrow \rangle$}}
\DeclareMathOperator*{\sgn}{sgn}
\newcommand{\be}{\begin{equation}}
\newcommand{\ee}{\end{equation}}

\newcommand{\beq}{\begin{eqnarray}}
\newcommand{\eeq}{\end{eqnarray}}

\def\H1{\widehat{H}_1}

\begin{document}
\noindent

\title{Quantum spin dynamics of mode-squeezed Luttinger liquids in two-component atomic gases}

\author{Artur Widera}\altaffiliation[Current address: ]{Institut f\"ur Angewandte Physik, Wegelerstr. 8, 53115 Bonn, Germany}\email{widera@uni-bonn.de}
\author{Stefan Trotzky}
\author{Patrick Cheinet}
\author{Simon F\"olling}
\author{Fabrice Gerbier}\altaffiliation[Current address: ]{Laboratoire Kastler Brossel, ENS, Universit\'e Pierre et Marie-Curie-Paris 6, CNRS ; 24 rue Lhomond, 75005 Paris, France}
\author{Immanuel Bloch}\affiliation{Johannes Gutenberg-Universit\"at, Institut f\"ur Physik, Staudingerweg 7,55099 Mainz, Germany}
\author{Vladimir Gritsev}
\author{Mikhail D. Lukin}
\author{Eugene Demler}\affiliation{Department of Physics, Harvard University, Cambridge MA 02138}

\begin{abstract}
We report on the observation of many-body spin dynamics of
interacting, one-dimensional (1D) ultracold bosonic gases with two
spin states. By controlling the non-linear atomic
interactions close to a Feshbach resonance we are able to induce a
phase diffusive many-body spin dynamics of the relative phase between the two components. We monitor this dynamical
evolution by Ramsey interferometry, supplemented by a novel,
many-body echo technique which unveils the role of quantum fluctuations in 1D. We find that the time evolution of the
system is well described by a Luttinger liquid initially prepared
in a multimode squeezed state. Our approach allows us to
probe the non-equilibrium evolution of one-dimensional
many-body quantum systems.

\end{abstract}

\maketitle
\begin{figure}
    \begin{center}
    \includegraphics[scale=0.9]{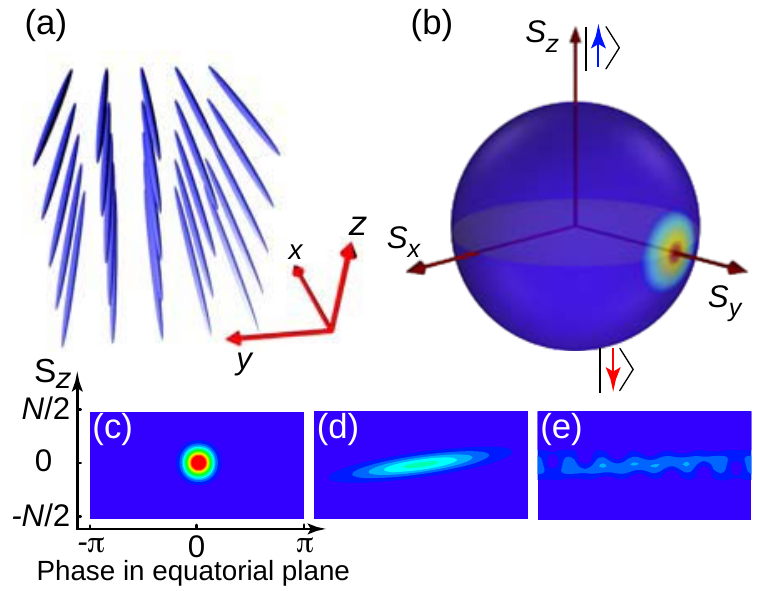}
    \end{center}
\caption{(a) Array of quasi-1D spinor systems. (b,c) Within each
tube, a coherent spin-state is created exhibiting Gaussian
distributed fluctuations of the mean spin. (c-e)
Time evolution of the initial CSS under non-linear interactions
re-distributing the initial Gaussian fluctuations towards increased
phase fluctuations assuming a SMA.\label{fig:Figure1}}
\end{figure}

 Among the applications of ultracold atomic gases, atom
interferometry stands out due to its potential for high precision
measurements \cite{Berman96}. In atom interferometry, the physical
quantity of interest is measured in terms of the relative phase
accumulated by the atomic wavefunction, subsequently mapped onto
atomic populations for efficient read-out. Due to their intrinsic
phase coherence and the possibility to create non-classical spin
states for precision metrology, Bose-Einstein condensates (BEC) seem
ideal candidates for such experiments. However, interatomic interactions mitigate this conclusion. For a
two-component interacting BEC, it has been shown
\cite{diffusion1,Sinatra99} using a single-mode approximation (SMA)
that the relative phase between the two components undergoes a
complicated evolution (Fig.~\ref{fig:Figure1},c-e), creating
quantum correlations \cite{Sorensen01} while single-particle coherence is
suppressed. Therefore, this dynamics is often termed {\em
phase diffusion}.

In this work, we investigate such an interaction induced dynamics in
quasi-1D two-component quantum gases by
monitoring the loss of coherence in a Ramsey-type
interferometer sequence. In order to distinguish different
contributions affecting the coherence
through the spin or spatial wave functions, we employ a novel
many-body spin echo sequence using a Feshbach resonance to adjust
sign and magnitude of the atomic interactions. When applied to a
single spatial mode BEC, this spin echo would lead to full revivals
of coherence, which are not observed in our experiment. In contrast,
quantum fluctuations play a key role for 1D interacting systems \cite{MW,Cornell},
which must necessarily be described as multimode quantum gases, as during the dynamical evolution higher energy modes become populated \cite{CH}.
The Luttinger liquid (LL) formalism \cite{GiamarchiBook,Caz}, which
reduces the interacting problem to an effective low-energy model of
decoupled harmonic oscillator modes, provides such a description. We
show theoretically that our preparation sequence amounts to
producing a multi-mode squeezed state in the \emph{spin
excitation modes}  of the LL oscillators, with each oscillator
itself prepared in a well-defined  mode-squeezed state, and
remaining in a squeezed state at all times \cite{footnote1}.
Monitoring the phase dynamics of this strongly non-equilibrium state
allows to probe fundamental aspects of 1D physics, namely the
competing dynamics of the (quasi-) condensate fraction (zero momentum
mode) and of the low-energy excitations, highly relevant for
squeezing experiments in 1D configurations \cite{Squeezing1D}. From
our model we find that only the lowest oscillator mode shows the
familiar revival dynamics, whereas the full model leads to the
partial revivals that we observe experimentally.

The system we consider is an array of two-component \Rb\ spinor
gases confined to quasi-1D traps (tubes). We experimentally realize this system by
loading a \Rb\ BEC of around $2.8\times10^5$ atoms into a 2D-optical
lattice \cite{Greiner01}, hence creating a 2D array of 1D degenerate
quantum gases (see Fig.~\ref{fig:Figure1}a). The lattice laser
wavelength is $\lambda=843\,$nm, and the radial and axial trap frequencies within each tubes are $\omega_r \approx 2\pi \times
42$\, kHz, and $\omega_{\mathit ax} \approx 2\pi \times 90\,$Hz,
respectively, from which we calculate a mean number per tube of $N=60$.

In order to extract information about the phase dynamics, we
experimentally monitor the coherence of the system by recording
interference fringes in a Ramsey-type interferometer. In the following we use the well known analogy with a quasi spin-1/2 system in order to describe our two-component gas. Starting from
a spin-polarized ensemble in state $\down \equiv \ket{F=1,m_F=+1}$,
we use a two-photon $\pi/2$-pulse combining a microwave and a radio
frequency photon to couple this state to the $\up = \ket{2,-1}$ spin
state and bring each atom into the single-particle superposition
$(\up+\down)/\sqrt{2}$. This prepares a coherent spin state (CSS)
within each tube with expectation value of the magnetization $\langle \hat{m}_z \rangle = 0$ and variance $\langle \hat{m}_z^2 \rangle = N/2$, with $\hat{m}_z = \hat
n_{\downarrow}-\hat n_{\uparrow}$ (see Fig.~\ref{fig:Figure1}b, cf.~\cite{ToBePublished07}). In order to observe interaction driven
effects, we let the system evolve for a given time at a particular
value of the inter-spin-state interaction strength, selected by using a Feshbach resonance
around $B=9.12\,$G. Thereby the
inter-species scattering length $a_{\uparrow\downarrow}$ can be changed by
a few 10\% from its background value \cite{vanKempen02,Widera04}, where $a_{ij}$ is the $s$-wave scattering length for
collisions between atoms in spin states $i$ and $j$.

After this time evolution, a final $\pi/2$-pulse with phase $\theta$
relative to the first pulse is applied, mapping the final relative
phase onto populations of spin states $\up$ and
$\down$ that are read out using state-selective absorption imaging.
In the absence of interactions and dephasing, such a sequence
results in sinusoidal Ramsey fringes in the relative population
$N_\uparrow / N_\mathrm{tot}$ as a function of $\theta$. Experimentally the coherence is
quantified through the visibility of the Ramsey fringe
\begin{equation}\label{eq:RamseyFringe}
     \frac{N_\uparrow}{N_\mathrm{tot}} = \frac{1}{2} \left( \vphantom{e^i} 1 + {\cal V}(t) \times \cos (\theta)
     \right),
\end{equation}
which is used to fit the experimental data and extract ${\cal V}(t)$
for a specific interaction time. Far from the Feshbach resonance
($B=8.7\,$G), where the effect of the phase dispersion can be
neglected, we measure a $e^{-1}$-decay time $t_\mathrm{dec}=54\,$ms,
which can be attributed to residual single particle decoherence
effects, e.g. caused by magnetic field
fluctuations.

\begin{figure}
    \centering
        \includegraphics[scale=0.85]{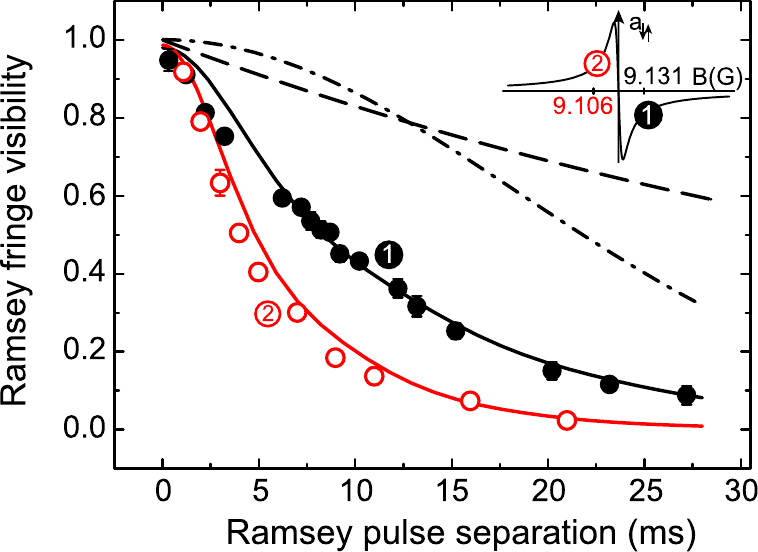}
    \caption{Ramsey fringe contrast drop for a time evolution at at $B=9.106\,$G ($\medcirc$) and $B=9.131\,$G ($\medbullet$).
    The dashed line indicates the independently measured decoherence far away from
    resonance ($t_\mathrm{dec}=54\,$ms).
    The solid lines are predictions of our LL model with the phase width
    $(\Delta \phi_{LL})^2 = (\Delta \phi_0)^2 \approx 0.033$ fixed. The dashed dotted line is a prediction of Eq.~(\ref{eq:VisibilityDecay}) based on SMA. For values see text.}
    \label{fig:SpinEcho_Figure2}
\end{figure}

Close to the Feshbach resonance, however, we find a markedly
faster decay of the Ramsey contrast. In
Fig.~\ref{fig:SpinEcho_Figure2} we monitor such a behavior of the
Ramsey fringe over time for two magnetic fields located almost
symmetrically around the center of the Feshbach resonance. Such a behavior can be expected from
enhanced phase diffusion due to increased interactions near the
resonance. Phase diffusion results from a spread in the distribution
of populations which are converted into phase fluctuations by the
non-linear interactions during the evolution, see
Fig.~\ref{fig:Figure1}c-e. In the simplest case where all atoms
occupy the same orbital wave function, the Ramsey fringe contrast
decays according to

\begin{equation}\label{eq:VisibilityDecay}
    {\cal V}_{\rm SMA}(t)\approx \exp \left( - \frac{1}{2}\chi^2 \langle \hat{m}_z^2 \rangle  t^2 \right).
\end{equation}

For the initial state we prepare, the population variance $\langle \hat{m}_z^2 \rangle = N/2$ leads to a phase uncertainty $(\Delta\phi_0)^2 =
2/N \approx 0.033$ of the collective spin vector in the equatorial
plane (see Fig.~1) in each tube, and a phase spreading time scale
$t_\phi \sim 1/(\chi \sqrt{N})$. The parameter $\chi$, related to
the second derivative of the chemical potential \cite{Sinatra99}, is
directly proportional to the difference $a_s = (a_{\uparrow\uparrow}
+ a_{\downarrow\downarrow} -2  a_{\uparrow\downarrow})/2$. Far from
the resonance, all three scattering lengths $a_{\uparrow\uparrow}$,
$a_{\downarrow\downarrow}$, and $a_{\uparrow\downarrow}$ are
approximately equal, so that $\chi\approx 0$ and interaction-induced
phase spreading can be neglected. However, near the Feshbach
resonance, the change of inter-species scattering length can lead to
a significant non-linear interaction energy. Following
Refs.~\cite{Petrov00,Paredes04,Sinatra99}, we estimate $\chi\approx
2 \pi \times 4.6\,$Hz for $B=9.131\,$G and our trapping parameters
with an atom number of $N \approx 60$. Although this value, together with the observed decoherence rate, is roughly on the order of the observed rate at which the coherence is lost in the quasi-1D
regime investigated here, the pure SMA Eq.~(\ref{eq:VisibilityDecay}) cannot explain our experimental observation in Fig.~\ref{fig:SpinEcho_Figure2}. Close to the  resonance we lose up to 50\% of the atoms due
to inelastic collisions. However, as these collisions usually remove
atoms from both spin states symmetrically, they do not modify the
magnetization of the system and thus only weakly influence the
dynamical evolution of the coherence for our measurement times
\cite{Sinatra99,footnote2}. 

In contrast to the simple model of Eq.~(\ref{eq:VisibilityDecay})
which predicts a symmetric decay around the resonance, we
systematically observe a faster drop of contrast below the
resonance. In fact, changing the sign of the effective interaction
strength has severe consequences on the spatial wave function of the
atoms. Below resonance, the inter-species repulsion is stronger
than the intra-species repulsion ($\chi<0$), and the system becomes
dynamically unstable towards demixing of the two species. This
reduces the Ramsey fringes visibility
below the resonance, which cannot be
distinguished from the effect of the coherent phase diffusion
dynamics.

\begin{figure}[htbp]
   \begin{center}
        \includegraphics[scale=0.75]{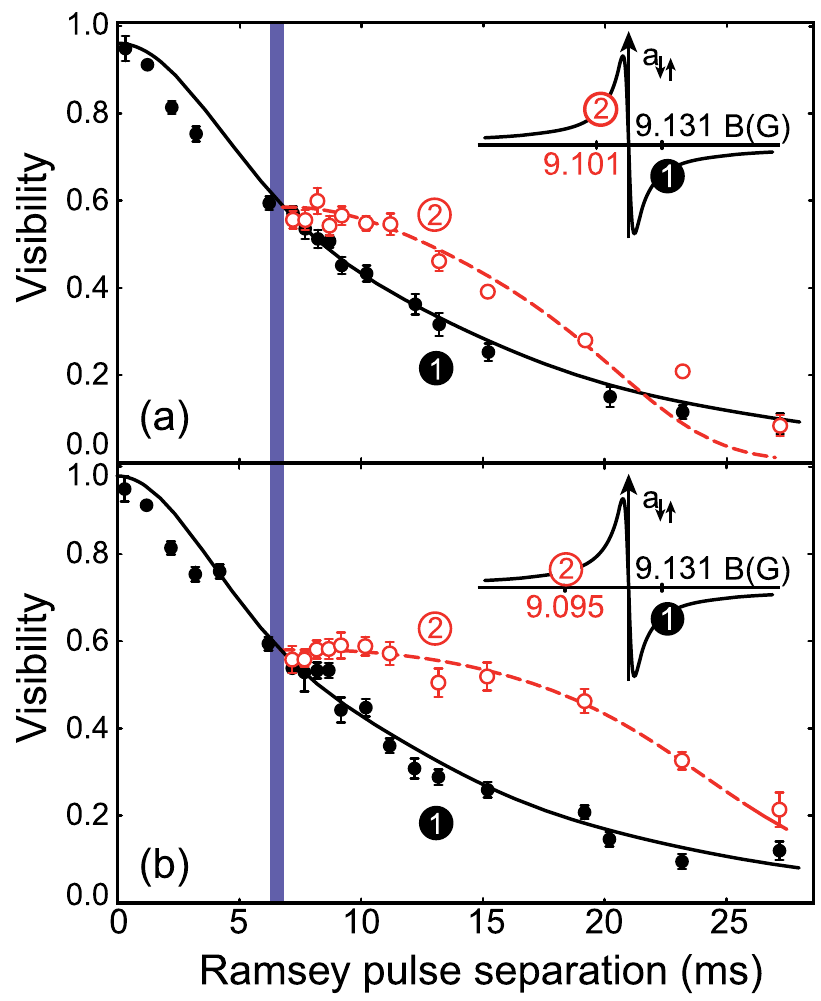}
    \end{center}
        \caption{Ramsey fringe visibility versus time for holding above the Feshbach resonance, $B=9.131\,$G ($\medbullet$), and with time reversal (spin echo) after 6.7\,ms ($\medcirc$) jumping from $B=9.131\,G$ to (a) $B=9.101\,$G and (b) $B=9.090\,$G.
        The solid and dashed lines are calculation from our model with $(\Delta \phi_{LL})^2=0.04$ (a), and $(\Delta \phi_{LL})^2=0.025$ (b).
        For $B=9.131\,G$, $a_s = 0.17\,a_{\uparrow\uparrow}$ and we compute $\gamma=0.165$ and
$K=8.0$.}
    \label{fig:Figure4new}
\end{figure}

In order to separate the effects of phase diffusion from other
mechanisms reducing the Ramsey fringe contrast, we apply a
many-body spin echo operation after an initial evolution time
$T$, similar to the one used in cavity quantum electro dynamics
experiments \cite{QEDSpinEcho}. We stress that our echo technique is acting on the \emph{many-body} quantum state, thereby extending previous theoretical work on echo operations neglecting phase diffusion \cite{Kuklov00}. Such a many-body spin echo operation is
performed by first holding the sample for a time $T=6\,$ms at a
magnetic field $B_1=9.131\,$G above the Feshbach resonance
($\chi>0$), and subsequently jumping below the resonance ($\chi<0$).
This operation effectively changes the sign of the non-linear
interaction parameter $\chi$, while heating or atom loss
can be avoided \cite{ToBePublished07}. In a SMA one would expect
this sequence to correspond to a perfect time reversal, leading to a
full revival of the contrast after another interaction time $T$
according to
\beq
\label{eq:SMAAfterSpinEcho}{\cal V}_{\rm SMA}(t)&=& {\cal V}_{0}\exp
\left( - \frac{\chi^2}{2} \langle \hat{m}_z^2 \rangle  (t-2T)^2
\right).
\eeq
This contradicts our observation of only partial revivals, shown for two spin echo sequences
in Fig.~\ref{fig:Figure4new}

In order to explain our observation, we model our system in a LL
approach going beyond the usual SMA \cite{diffusion1,CH}. A drastic
simplification follows from the near equality
$a_{\downarrow\downarrow}\approx a_{\uparrow\uparrow}$, which
results in a decoupling of elementary excitations into almost
independent density and spin fluctuations. The latter can be
described in terms of two conjugate fields, $\hat m_{z}$ and $\hat \phi_{s}$, describing
respectively fluctuations of the local magnetization and of the
relative phase. At long-wavelengths, the spin-part of the
Hamiltonian reads
\begin{equation}
H_s=\int dx\, \left [ g_s \hat m_{z}^2 + \frac{n_{\text {tot}}}{4M}
\left( \nabla\hat \phi_s\right)^2 \right],\label{eq:SpinHamiltonian}
\end{equation}
with $n_{\text{tot}}=n_\uparrow+n_\downarrow$ the linear
density. For a uniform 1D system of length $L$, the fields $\hat
m_z$ and $\hat\phi_s$ can be expanded in terms of momentum
eigenmodes $\hat{a}_q,\hat{a}_{q}^{\dag}$ as \cite{Haldane}
$\hat{\phi}_{s}(x) 
=\phi_{0}+\sum_{q}(2qLK/\pi)^{-1/2}
e^{-\vert q \vert /2q_c} \sgn(q) \left[ e ^{i q x}
\hat{a}_q^{\phantom\dagger}+h.c. \right ]$.
Each mode is characterized by the wave vector $q$ and frequency
$\omega_{q}(t)=v_{s}(t)|q|$, where $v_{s}(t)=(g_{s}(t)
n_{\text{tot}}/M)^{1/2}$ is the spin velocity and $g_s(t)$ denotes the
spin coupling constant. The sum over LL modes exclude the zero mode,
and are restricted to values of $q$ below a cut-off momentum
$q_c\sim\xi_h^{-1}$, where $\xi_h$ is the healing length. The
interactions are encoded in the LL parameter $K$ which can take
values ranging from 1 (the so-called Tonks-Girardeau limit
\cite{Paredes04,Kinoshita04}) to infinity (noninteracting gas). For
weakly interacting bosons \cite{Caz} $K \approx
\pi/[\sqrt{\gamma}(1-\sqrt{\gamma}/2\pi)^{1/2}]$ where
$\gamma=2a_{s} M \omega_{r}/\hbar n_\mathrm{tot}$. In our
experiments \cite{Widera04} $\gamma\sim 0.1-0.2$  and $K\sim 5-8$
for different data sets. The contribution of the zero energy mode
$\phi_{0}$ is identical to that in the SMA, as described before
\cite{diffusion1}. The dynamics of the low-energy LL excitations on
top of the zero mode dynamics is that of a collection of independent
harmonic oscillators with (time-dependent) frequencies $\omega_{q}$.

In order to use the LL decomposition to compute the time evolution,
we need to identify how to describe the initial state in terms of
those LL modes. Our experimental scheme ideally corresponds to an
instantaneous projection of the spin state (initially polarized in
$\down$) onto a state with zero relative phase directly after the
first $\pi/2$-pulse, $\hat{\phi}_{s}(x)\ket{\psi(0)} \approx 0$. The
connection with the LL formalism is done by identifying the initial
CSS as a multimode squeezed state for the elementary spin
excitations,
\begin{eqnarray}
|\psi(0)\rangle=\hat{S}(\{w_{q}\})|0\rangle
\!=\!\prod_{q}\sqrt{\left(1-|w_{q}|^{2}\right)}
\exp(w^{\vphantom{\dag}}_{q}\hat{a}^{\dag}_{q}\hat{a}^{\dag}_{-q})|0\rangle
\end{eqnarray}
where $\hat{S}$ is a squeezing operator, with the factor
$w_{q}=(1-\alpha_{q})/(1+\alpha_{q})$. Here, $\alpha_{q}=\Delta
\phi_{LL} |q|/|q_{c}|$ is a mode squeezing parameter, and 
the phase variance $(\Delta\phi_{LL})^2=(\Delta\phi_{0})^2=\frac{2}{N}$. In
reality, due to various experimental imperfections (e.g.~unknown
temperature), the exact width of the prepared squeezed state is
unknown and cannot be determined independently. We
still consider the initial state as a squeezed state of the LL
oscillators, with a fitted $\Delta \phi_{LL}$ of the squeezed state.

The time evolution of a squeezed state under the LL Hamiltonian
amounts to the replacement $\hat{a}_q\rightarrow
\hat{a}_q\exp(-i\omega_{q}t)$. In addition, the reversal of the sign
of interaction at time $t=T$ amounts to a sign reversal of
the spring constant of each LL harmonic oscillator. We are able to
compute this time evolution exactly \cite{ToBePublished07}, using the formalism of harmonic
oscillators with time-dependent frequencies $\omega_{q}(t)$
\cite{osc}. Here, we concentrate on
the comparison between the predictions of the calculations and the
experimental results. 
From our model we are able to calculate the coherence factor ${\cal V}(t) = \mathit{Re}\left\{
\frac{1}{L}\int_0^L \,dx\, \bra{\psi(t)} e^{i\,\hat{\phi}_s}
\ket{\psi(t)} \right\}$, measuring the relative phase $\hat \phi_s$ 
between $|\!\!\uparrow\rangle$ and $|\!\!\downarrow\rangle$.
In the LL formalism, the coherence factor can
be generally written as ${\cal V}(t)={\cal V}_\mathrm{SMA}(t)\times
{\cal V}_{q\neq 0}(t)$, where ${\cal V}_\mathrm{SMA}(t)$ is given by
Eqs.~(\ref{eq:VisibilityDecay},\ref{eq:SMAAfterSpinEcho}), and the
term describing the contribution of the $q\neq 0$ modes to the decay
of contrast ${\cal V}_{q\neq 0}(t)$ is known explicitly
\cite{ToBePublished07}. The combination of these effects leads to
the typical behavior of ${\cal V}(t)$ illustrated in
Fig.~\ref{fig:Figure4new} where we quantitatively compare the
experimental results with the predictions of our LL model. The interaction parameters of the LL model were
determined from the microscopic data \cite{Widera04}, and in our
computations we only consider the density in the central tubes
computed in the Thomas-Fermi approximation \cite{footnote2}. Overall, we find very good
agreement between the LL model and the experimental data using
$\Delta \phi_{LL}$ as the only fit parameter. For our data we find values that are of the same order of magnitude as the initial width.
For longer times ($t>20\,$ms), the model deviates from the measured data. This
breakdown is due to the phenomenon of demixing discussed above, when the excitations become so strong that density fluctuations are significant \cite{Cornell}.
Its timescale can be estimated as the time required for the
formation of random magnetization domains, when $\sum_{q}\langle
|\hat{m}_{q}|^{2}\rangle\sim N$. For our experimental parameters,
it is of the order $\approx 20-25$\,ms which is of the order of
$1/\chi$.

From this analysis we find indeed that although the zero mode
evolution is perfectly refocused by changing the sign of $g_s$, the
non-zero modes still undergo dephasing even under the echo sequence.
The reason for this is the kinetic energy term in Eq.~(\ref{eq:SpinHamiltonian}), unaffected by the spin echo.
Hence, the reversal is exact for the
$q=0$ mode, significant for low-lying spin waves with $q\sim 1/L$,
but increasingly less efficient for higher lying spin waves modes
with $L^{-1}\ll q < q_c$. Note that a full revival could be in
principle achieved by also reverting the second, kinetic energy term
of the Hamiltonian in Eq.~(\ref{eq:SpinHamiltonian}). This could be
realized by inducing a negative effective mass, {\it e.g.} through a
weak optical lattice along the direction of the tubes
\cite{Oberthaler03}.

In conclusion, we have studied the phase dynamics of quasi-1D
two-component quantum gases with adjustable interaction. For strong
interactions we observe an accelerated decay of coherence in the
system. By application of a novel many-body spin echo technique we
are able to reverse the interaction driven dynamics, leading to a
partial revival of the coherence in the system. We attribute this
revival to the dynamics of the ground state mode, reflecting the
coherent nature of the phase diffusion dynamics. This is supported
by quantitative comparison to our LL model. The missing fraction of
coherence in the revival is also quantitatively in agreement with
our model and demonstrates the importance of the dynamical evolution
of higher lying modes in 1D systems. Our experiment shows that
quantum fluctuations are a crucial component in the discussion of
phase diffusion dynamics \cite{diffusion1} and spin-squeezing
\cite{Sorensen01,Squeezing1D} in low dimensional systems. While our
work demonstrates that these quantum fluctuations fundamentally
limit the performance of atom interferometers in 1D, it also
indicates an avenue to overcome such limitations by inverting both
interaction and kinetic energy terms simultaneously during the
interferometer sequence.

We thank Servaas Kokkelmans for providing us with updated
calculations on the hyperfine Feshbach resonance. We acknowledge
financial support by the DFG, and the EU under STREP (OLAQUI) and
MC-EXT (QUASICOMBS). FG acknowledges support from IFRAF and ANR. VG,
ED and ML acknowledge support from NSF, Harvard-MIT CUA, AFOSR, and
MURI. VG is also supported by Swiss NSF.


\begin{thebibliography}{20}
\bibitem{footnote1}
We note that this mode-squeezing is different from squeezing the $\hat{S}_z$ spin-component during phase diffusion.
\bibitem{footnote2}
The dependency of the interaction strength $\chi$ on density is weak. Therefore we assume a constant interaction strength $\chi$ over time, even for the cases close to the resonance.
\end{thebibliography}
\end{document}